\documentclass[aps,pra,twocolumn,superscriptaddress,floatfix,nofootinbib]{revtex4-2}
\usepackage{amsmath,amssymb,amsthm,bm,braket,graphicx}
\usepackage[colorlinks=true,citecolor=blue,linkcolor=blue,urlcolor=blue]{hyperref}

\newtheorem{theorem}{Theorem}
\newtheorem{proposition}{Proposition}
\newtheorem{corollary}{Corollary}
\theoremstyle{definition}
\newtheorem{definition}{Definition}

\newcommand{\tr}{\operatorname{tr}}
\newcommand{\diag}{\operatorname{diag}}
\newcommand{\id}{\openone}
\newcommand{\Hs}{\mathcal H}
\newcommand{\Free}{\mathcal F}
\newcommand{\cpp}[1]{\ket{c_{#1}^{\perp}}}
\newcommand{\ketbra}[2]{|#1\rangle\langle #2|}
\newcommand{\Ml}{\mathcal M_{\ell_1}}
\newcommand{\Ct}{\widetilde C}
\newcommand{\EC}{E_C}
\newcommand{\El}{E_{\ell_1}}
\newcommand{\Cmax}{C_{\max}(d)}

\begin{document}

\title{Quantifying superposition via coherence by stochastic orthogonalization}

\author{G\"okhan Torun}
\email{gtorung@gmail.com}
\affiliation{Department of Physics, Faculty of Arts and Sciences, \.{I}zmir University of Economics, 35330 \.{I}zmir, T\"urkiye}

\date{\today}

\begin{abstract}
Here we present an approach to quantifying superposition of nonorthogonal reference states through quantum coherence. No change of basis achieves this: every orthonormal basis makes some superposition-free states appear coherent. We show that orthogonalization succeeds instead as a measurement mapping every reference state onto a multiple of a basis vector in each outcome; such measurements are exactly the free operations that convert superposition into coherence. The largest average coherence they extract from a state, which we call its degree of superposition, turns every coherence measure into a superposition measure with the same operational meaning. An interference term becomes coherence at most with the probability of unambiguously discriminating its two reference states, a bound attained when there are no others and, for positive coefficients and nonnegative overlaps, exactly when the overlaps are probabilities of a random grouping. Maximal degree means conversion into a maximally coherent state with certainty, possible for up to three reference states but not in general. Each coherence measure gives its own degree; when two degrees order two superpositions oppositely, neither converts into the other with certainty, and we find such pairs already for two reference states, unlike in qubit coherence. For equal positive overlaps, in-phase superpositions are rigid: deterministic conversions that keep all components can only permute their coefficients. As the reference states merge, the degree stays bounded where widely used measures can diverge, and at every amplitude a rotation-symmetric cat state attains its maximum. Our results quantify superposition operationally and reveal conversion constraints created by nonorthogonality.
\end{abstract}

\maketitle

\section{Introduction}
Quantum coherence, the superposition of the states of a fixed orthonormal basis, is the best understood form of superposition as a resource. In the resource theory of coherence, the free states are the incoherent ones; the free operations cannot create coherence from them; and a large family of measures quantifies it~\cite{Aberg2006,Baumgratz2014,Winter2016,Streltsov2017,ChitambarGour2019}. Many superpositions of physical interest, however, are built from nonorthogonal states, including coherent states of light, as in cat states~\cite{Dodonov1974,Ourjoumtsev2006}, spin-coherent states~\cite{Arecchi1972}, and the atomic orbitals of quantum chemistry~\cite{Lowdin1950,Chirgwin1950,Gallup2002}. The resource theory of superposition (RTS)~\cite{Theurer2017,Killoran2016} treats such superpositions as a resource, with the nonorthogonal states as reference states. State transformations, golden states, majorization, and distillation in RTS have been characterized~\cite{Torun2021,Senyasa2022,Torun2026Gram,Torun2026Distill}, and a related framework describes coherence with respect to general measurements~\cite{Bischof2019}. Yet, unlike coherence theory, RTS has few measures with closed forms and clear operational meanings~\cite{Theurer2017,Torun2026Gram}.

The problem addressed here is how superposition can be measured through coherence, so that the measures of coherence theory and their operational meanings become available in RTS. A direct approach orthogonalizes the reference states into a basis of the same space and evaluates a coherence measure in that basis. This cannot work: in no orthonormal basis are two nonorthogonal reference states both incoherent (Proposition~\ref{prop:nogo}), so in every such basis some superposition-free states appear coherent. Tan \emph{et al.}~\cite{Tan2017} used an orthogonalization procedure to quantify the coherence between coherent states within a resource theory of linear optics; in RTS, by contrast, only a fixed finite set of reference states and their mixtures are free~\cite{Theurer2017}.

Our approach keeps the purpose of orthogonalization, that superposition-free states become incoherent, but replaces the change of basis by a measurement each of whose outcomes maps every reference state onto a multiple of a basis vector. These stochastic orthogonalizations are exactly the free operations that convert superposition into coherence; in Ref.~\cite{Torun2026Distill} such maps gave the optimal probability of reaching a maximally coherent state. Here we ask how much coherence they extract on average and define the degree of superposition of a state as the largest average coherence into which it can be converted. Quantifying one resource by the amount of another that can be obtained from it is an established strategy, used for nonclassical light~\cite{Asboth2005,Killoran2016}, quantum discord~\cite{Streltsov2011,Piani2011}, and coherence~\cite{Streltsov2015,ChitambarHsieh2016,Ma2016,Regula2018}; there the target resource is a correlation between subsystems, whereas here it is the coherence of a single system. Since a conversion succeeds only as far as the reference states can be told apart, unambiguous discrimination~\cite{Ivanovic1987,Dieks1988,Peres1988,Chefles1998} and the interference--distinguishability tradeoff of multipath interferometry~\cite{Englert1996,Bera2015,Bagan2016} enter the measures directly.

Our results are as follows. First, for every coherence measure $C$ the degree of superposition $\EC$ is a superposition measure that reduces to $C$ for orthonormal reference states (Theorem~\ref{thm:measure}). Second, the states that can be converted with certainty into a maximally coherent state reach the largest possible value of $\EC$ for every $C$, and for the $\ell_1$-norm and the relative entropy of coherence they are the only ones (Theorem~\ref{thm:max}). These maximal-degree states exist whenever there are at most three reference states, but not in general when there are more; when they exist, they are exactly the states whose coefficients have the largest product of squared moduli, and one unambiguous discrimination converts them all (Theorem~\ref{thm:exist}). Third, when there are only two reference states, unambiguous discriminations are optimal for every coherence measure, and for the $\ell_1$-norm of coherence $\El=(1-|s|)\Ml$, the $\ell_1$-measure of superposition~\cite{Theurer2017} multiplied by the optimal probability $1-|s|$ of unambiguously discriminating two reference states with overlap $s$ (Theorem~\ref{thm:qubit}).

Fourth, the degrees constrain transformations: each of them bounds the probability of a conversion, and since each coherence measure gives its own degree, two degrees can order two superpositions oppositely, so that neither converts into the other with certainty. $\El$ and the degree for the relative entropy of coherence reveal such incomparable superpositions already for two reference states, whereas all coherence measures order the pure states of a qubit alike. A deterministic conversion that keeps all components, moreover, never lowers the interference term of the norm of a pure state~\cite{Torun2026Distill}, and for equal positive overlaps and any number of reference states, constructively interfering superpositions are rigid, since such conversions can only permute their coefficients, and form continuous families of pairwise incomparable states (Proposition~\ref{prop:rigidd}). For two reference states, the interference term and the imbalance of the weights decide every deterministic conversion with real coefficients (Proposition~\ref{prop:rigid}).

Fifth, for any number of reference states, the interference term $M_{jk}$ between two components is converted into coherence with weight at most $1-|S_{jk}|$, where $S_{jk}$ is the overlap of the two reference states and $1-|S_{jk}|$ the optimal probability of discriminating them unambiguously (Theorem~\ref{thm:bounds}). For superpositions with positive coefficients of reference states with nonnegative overlaps, this pairwise bound is attained iff each overlap is the probability that a random grouping of the reference states puts the two states into the same class (Theorem~\ref{thm:grouping}, Proposition~\ref{prop:mixture}); for uniform overlaps it is attained for every state (Corollary~\ref{cor:sym}). The bound is itself a superposition measure in closed form, one of a family weighted by distinguishability (Theorem~\ref{thm:weighted}). The degree depends on the whole reference set: adding reference states never increases it, and can decrease it strictly (Sec.~\ref{sec:general}). Finally, we treat two-component cat states, three coherent states on a line, and the rotation-symmetric cat states of bosonic codes~\cite{Mirrahimi2014,Grimsmo2020} (Sec.~\ref{sec:examples}); there the degree remains bounded, whereas the $\ell_1$-measure and the robustness of superposition~\cite{Theurer2017} can diverge as the coherent states approach each other.

The paper is organized as follows. Section~\ref{sec:setting} fixes the setting, Sec.~\ref{sec:orth} defines stochastic orthogonalization, and Sec.~\ref{sec:measure} constructs the measures and characterizes the maximal-degree states. Sections~\ref{sec:two}--\ref{sec:weighted} treat two reference states, incomparable superpositions, any number of reference states, and the weighted measures; Sec.~\ref{sec:examples} gives the examples, and Sec.~\ref{sec:conclusion} concludes.

\section{Setting}\label{sec:setting}
Let $\ket{c_0},\dots,\ket{c_{d-1}}$ be normalized, linearly independent reference states that span $\Hs\cong\mathbb C^d$, with the positive definite Gram matrix $S$, $S_{jk}=\braket{c_j|c_k}$. The reference states thus form a basis of $\Hs$, in general nonorthogonal, and $d=\dim\Hs$ is their number; a statement about two or three reference states concerns $d=2$ or $d=3$, not a state that superposes two or three states of a larger reference set. A bar denotes complex conjugation and a dagger the adjoint; inequalities between Hermitian matrices refer to the positive semidefinite order. With an orthonormal basis $\{\ket j\}$ of a label space $\mathbb C^d$ we write $V=\sum_j\ketbra{c_j}{j}$, so that $V^\dagger V=S$; $V$ is invertible because the reference states are linearly independent and span $\Hs$. The dual vectors $\cpp j$ are defined by $\braket{c_j^\perp|c_k}=\delta_{jk}$. Every pure state is $\ket\psi=\sum_jx_j\ket{c_j}$ with a unique coefficient vector $x$, $x_j=\braket{c_j^\perp|\psi}$, normalized as $x^\dagger Sx=1$, and every state has a unique coefficient matrix,
\begin{equation}
\rho=\sum_{jk}M_{jk}\ketbra{c_j}{c_k},\quad M=V^{-1}\rho\,(V^{-1})^\dagger\ge0, \label{eq:M}
\end{equation}
and the normalization $\tr\rho=1$ reads $\tr(SM)=1$. The free states are the mixtures of reference states; they form the set $\Free$, and a state is free iff $M$ is diagonal~\cite{Theurer2017}. A Kraus operator is superposition free if it sends each reference state to a multiple of a reference state; on coefficient vectors it acts as the matrix $\hat K=\Pi_hD_\xi$, where $(\Pi_h)_{jk}=\delta_{j,h(k)}$ for a map $h$ of the index set into itself and $D_\xi=\diag(\xi)$. A channel is a superposition-free operation (FO)~\cite{Theurer2017} if it has a Kraus decomposition into such operators, whose completeness reads~\cite{Torun2026Distill}
\begin{equation}
\sum_n\hat K_n^\dagger S\hat K_n=S. \label{eq:FO}
\end{equation}
An FO maps free states to free states, so it cannot create superposition. A function $\mathcal M$ on states is a superposition measure if it is (M1) faithful, $\mathcal M(\rho)\ge0$ with equality iff $\rho\in\Free$; (M2) monotonic, $\mathcal M(\Lambda(\rho))\le\mathcal M(\rho)$ for every FO $\Lambda$; (M3) strongly monotonic, $\sum_np_n\mathcal M(\rho_n)\le\mathcal M(\rho)$ for the outcomes $\rho_n=K_n\rho K_n^\dagger/p_n$ of every FO, with $p_n=\tr(K_n\rho K_n^\dagger)$; and (M4) convex~\cite{Theurer2017}. The off-diagonal entries $M_{jk}$, $j\ne k$, are the interference terms between the components $j$ and $k$; the $\ell_1$-measure of superposition is their total modulus, $\Ml(\rho)=\sum_{j\ne k}|M_{jk}|$~\cite{Theurer2017}. A state is in-phase if $M_{jk}\ge0$ for all $j,k$; a pure state is in-phase iff its nonzero coefficients have a common phase. We also use $\nu(\rho)=\tr M$, which for a pure state is the squared norm $\sum_j|x_j|^2$ of its coefficient vector, and the coefficient state $\sigma_\rho=M/\tr M$. A pure state $\ket\psi$ is also denoted by $\psi$, as in $\nu(\psi)$, and likewise for $\Psi_\pm$, $\Phi_m$, and $\chi$ below.

Coherence refers to a label space $\mathbb C^N$ with a fixed orthonormal basis $\{\ket\mu\}$, $\mu=0,\dots,N-1$, for any $N$. It is the special case of RTS in which the reference states are orthonormal: incoherent states are diagonal, and incoherent Kraus operators send each basis vector to a multiple of a basis vector~\cite{Baumgratz2014,Winter2016}. By a coherence measure $C$ we mean a function on the states of all label spaces that is faithful, convex, and strongly monotonic under incoherent instruments between label spaces of any dimensions; examples are the $\ell_1$-norm of coherence $C_{\ell_1}(\tau)=\sum_{\mu\ne\mu'}|\tau_{\mu\mu'}|$, the relative entropy of coherence $C_{\rm r}$, and the robustness of coherence~\cite{Baumgratz2014,Napoli2016}. We write $\Cmax$ for the largest value of $C$ on states supported on at most $d$ basis vectors; it is attained by the maximally coherent states~\cite{Baumgratz2014}, which are $\sum_{\mu=0}^{d-1}\ket\mu/\sqrt d$ up to phases and a relabeling of the basis, and $C_{\ell_1}$ and $C_{\rm r}$ attain it only on these states. For the $\ell_1$-norm of coherence, $C_{\ell_1,\max}(d)=d-1$. We extend $C$ to positive semidefinite matrices by $\Ct(Y)=\tr(Y)\,C(Y/\tr Y)$ and $\Ct(0)=0$. This extension is positively homogeneous and convex, and the average coherence of an ensemble $\{p_n,\tau_n\}$ is $\sum_n\Ct(p_n\tau_n)$.

Coherence and superposition refer to different structures. Coherence is defined relative to an orthonormal basis, and superposition relative to the reference states: a state is a superposition state if it is not a mixture of reference states. The maximally mixed state $\id/N$ of a label space is incoherent. The maximally mixed state $\id/d$ of $\Hs$ is diagonal in every orthonormal basis of $\Hs$, yet it is a superposition state unless $S=\id$, because its coefficient matrix $S^{-1}/d$ is not diagonal~\cite{Torun2026Gram}; the uniform mixture $\sum_j\ketbra{c_j}{c_j}/d$ of the reference states is superposition free.

We now fix terms for the counterparts of maximally coherent states in RTS, which are compared throughout the paper. In coherence theory every state can be prepared from a maximally coherent state by incoherent operations~\cite{Baumgratz2014}, so the maximally coherent states maximize every coherence measure. Their counterparts in RTS, the golden states, from which every state can be prepared deterministically by FO, exist for $d\ge3$ only for special Gram matrices~\cite{Theurer2017,Senyasa2022,Torun2026Gram}; without them, different superposition measures are in general maximized by different states. We therefore distinguish four kinds of states. A golden state is a state with maximal superposition in the sense of Ref.~\cite{Theurer2017} and a root state in the sense of Ref.~\cite{LiuBuTakagi2019}, and by (M2) it maximizes every superposition measure. Maximally superposed states maximize the robustness and the geometric measure of superposition; they exist for every $S$, and their overlaps $\braket{c_j|\psi}$ have equal moduli~\cite{Torun2026Gram}. Maximal-degree states (Sec.~\ref{sec:measure}) maximize the degree of superposition introduced here; they exist for $d\le3$, but not for every $S$ with $d\ge4$. Balanced states, whose coefficients $x_j$ have equal moduli~\cite{Torun2026Distill}, are defined by their form rather than by a maximum. A golden state is maximally superposed, of maximal degree, and balanced (Sec.~\ref{sec:measure} and Ref.~\cite{Torun2026Gram}), but a maximal-degree state need not be golden, maximally superposed, or balanced (Sec.~\ref{sec:examples}).

\section{Orthogonalization as a measurement}\label{sec:orth}
\subsection{No orthogonalization by a change of basis}
An orthogonalization in the usual sense replaces the reference states by an orthonormal basis $\ket{e_j}=\sum_kT_{kj}\ket{c_k}$ of $\Hs$, with $T^\dagger ST=\id$. The Gram--Schmidt procedure corresponds to a triangular $T$, and the symmetric (L\"owdin) orthogonalization~\cite{Lowdin1950} to $T=S^{-1/2}$; the latter has been used to relate coherence and superposition~\cite{Torun2023}. One then measures the coherence of $\rho$ in the basis $\{\ket{e_j}\}$.

\begin{proposition}[No deterministic orthogonalization]\label{prop:nogo}
Let the reference states be embedded isometrically into a space with an orthonormal basis. Any two reference states that are not orthogonal are not both incoherent in that basis. Hence, unless $S=\id$, no orthonormal basis makes all free states incoherent, and the coherence in any orthonormal basis violates (M1).
\end{proposition}
\begin{proof}
A pure state is incoherent iff it is a basis vector up to a phase. Two basis vectors are orthogonal or equal up to a phase, whereas distinct reference states are linearly independent.
\end{proof}

\subsection{Stochastic orthogonalization}
The purpose of an orthogonalization can still be achieved if the map is allowed to have several outcomes.

\begin{definition}[Stochastic orthogonalization]\label{def:orth}
A stochastic orthogonalization is an instrument $\{A_n\}$ from $\Hs$ to a label space $\mathbb C^N$, $\sum_nA_n^\dagger A_n=\id$, each of whose Kraus operators sends every reference state to a multiple of a basis vector,
\begin{equation}
A_n\ket{c_j}=\xi_{n,j}\ket{h_n(j)}, \label{eq:A}
\end{equation}
where $h_n$ maps the indices of the reference states to labels of the basis.
\end{definition}

Since coherence theory is RTS with orthonormal reference states, Eq.~\eqref{eq:A} is the condition for a free operation of RTS from the reference states $\{\ket{c_j}\}$ to the basis $\{\ket\mu\}$. Stochastic orthogonalizations are therefore exactly the free operations that convert superposition into coherence: they cannot create coherence from a superposition-free state. On coefficient matrices, outcome $n$ acts by an incoherent Kraus operator,
\begin{equation}
\begin{aligned}
\hat A_n&=A_nV=\Pi_{h_n}D_{\xi_n},\\
p_n&=\tr\big(\hat A_nM\hat A_n^\dagger\big),\quad \tau_n=\hat A_nM\hat A_n^\dagger/p_n,
\end{aligned} \label{eq:Ahat}
\end{equation}
where $(\Pi_{h_n})_{\mu j}=\delta_{\mu,h_n(j)}$, $p_n$ is the probability of the outcome, and $\tau_n$ is the resulting state of the label space. What distinguishes a stochastic orthogonalization from an incoherent operation is its normalization.

\begin{proposition}[Decompositions of the Gram matrix]\label{prop:decomp}
(a) The operators~\eqref{eq:A} form a stochastic orthogonalization iff
\begin{equation}
\begin{gathered}
\sum_n\hat A_n^\dagger\hat A_n=S,\quad\text{that is,}\\
\sum_{n:\,h_n(j)=h_n(k)}\bar\xi_{n,j}\,\xi_{n,k}=S_{jk}\quad\text{for all }j,k.
\end{gathered} \label{eq:decomp}
\end{equation}

(b) In particular, the squared amplitudes of every reference state sum to one,
\begin{equation}
\sum_n|\xi_{n,j}|^2=1\quad\text{for all }j, \label{eq:norm}
\end{equation}
and the overlap $S_{jk}$ is carried entirely by the outcomes that map $\ket{c_j}$ and $\ket{c_k}$ onto the same basis vector.
\end{proposition}
\begin{proof}
(a) Since $V$ is invertible, $\sum_nA_n^\dagger A_n=\id$ is equivalent to $V^\dagger\big(\sum_nA_n^\dagger A_n\big)V=S$, and Eq.~\eqref{eq:A} gives
\begin{equation}
\big(\hat A_n^\dagger\hat A_n\big)_{jk}=\bar\xi_{n,j}\,\xi_{n,k}\,\delta_{h_n(j),h_n(k)}. \label{eq:entries}
\end{equation}

(b) Equation~\eqref{eq:norm} is the diagonal of Eq.~\eqref{eq:decomp}, and by Eq.~\eqref{eq:entries} only outcomes with $h_n(j)=h_n(k)$ contribute to $S_{jk}$.
\end{proof}

Each term $\hat A_n^\dagger\hat A_n$ is block diagonal with respect to the classes of reference states that $h_n$ maps onto the same basis vector, and each of its blocks has rank one. Stochastic orthogonalizations are thus the decompositions of the Gram matrix into such matrices, whereas the symmetric and the Gram--Schmidt orthogonalization correspond to the single-term factorizations $S=S^{1/2}S^{1/2}$ and $S=R^\dagger R$ with triangular $R$, which are not of this form. A single outcome would require $|\xi_j|=1$ by Eq.~\eqref{eq:norm}, hence $|S_{jk}|\in\{0,1\}$, so every stochastic orthogonalization has several outcomes unless $S=\id$.

Two kinds of outcomes recur below. The first discriminates the reference states unambiguously. Let $a=(a_0,\dots,a_{d-1})$ with $a_j\ge0$ and $S-D_a\ge0$, where $D_a=\diag(a)$. The filter $\hat A_0=D_a^{1/2}$ together with operators $\hat A_l=\ketbra{f}{b_l}$, $l=1,2,\dots$, with $\sum_l\ketbra{b_l}{b_l}=S-D_a$, which map all reference states onto an additional basis vector $\ket f$, satisfies Eq.~\eqref{eq:decomp}. The filter identifies $\ket{c_j}$ with probability $a_j$, and the outcome $f$ is inconclusive. The condition $S-D_a\ge0$ describes the success probabilities attainable in unambiguous discrimination~\cite{Chefles1998}; equal probabilities are possible up to $a_j=\lambda_{\min}(S)$, and for $d=2$, with overlap $s$, the condition reads $(1-a_0)(1-a_1)\ge|s|^2$~\cite{JaegerShimony1995}, with the unbiased optimum $a_0=a_1=1-|s|$~\cite{Ivanovic1987,Dieks1988,Peres1988}.

The second groups the reference states. For a partition $P$ of the index set into classes, let $\Pi_P$ map each $\ket j$ onto the basis vector of its class, and let $B_P=\Pi_P^\dagger\Pi_P$, that is, $(B_P)_{jk}=1$ if $j$ and $k$ lie in the same class and $0$ otherwise. An outcome $\sqrt{w_P}\,\Pi_P$, with a weight $w_P>0$, identifies the class of a reference state but not the state within its class. It discards the interference terms within the classes and converts those between classes into coherence, $(\Pi_PM\Pi_P^\dagger)_{\mu\mu'}=\sum_{j\in\mu,k\in\mu'}M_{jk}$ for classes $\mu$ and $\mu'$.

\section{Superposition measures from coherence measures}\label{sec:measure}
A stochastic orthogonalization turns $\rho$ into an ensemble $\{p_n,\tau_n\}$ of states of coherence theory, and since it is a free operation, the coherence of this ensemble is extracted from the superposition of $\rho$.

\begin{definition}[Degree of superposition]\label{def:degree}
Let $C$ be a coherence measure. The degree of superposition of a state $\rho$ with respect to $C$ is the largest average coherence into which $\rho$ can be converted by a stochastic orthogonalization,
\begin{equation}
\EC(\rho)=\sup_{\{A_n\}}\sum_np_n\,C(\tau_n)=\sup_{\{A_n\}}\sum_n\Ct\big(\hat A_nM\hat A_n^\dagger\big), \label{eq:E}
\end{equation}
where the supremum is over all stochastic orthogonalizations, with any number of outcomes and any label space.
\end{definition}

We write $\El$ for the degree of superposition with respect to the $\ell_1$-norm of coherence. The degree of superposition is thus measured in the units of coherence: a state has a larger degree than another if free operations extract more coherence from it on average.

\begin{theorem}[Superposition measures]\label{thm:measure}
Let $C$ be a coherence measure. Then, for every set of reference states,

(a) $\EC$ is a superposition measure: it satisfies (M1)--(M4);

(b) $\EC=C$ for orthonormal reference states, under the identification $\ket{c_j}=\ket j$;

(c) for every state $\rho$,
\begin{equation}
\lambda_{\min}(S)\,\nu(\rho)\,C(\sigma_\rho)\le\EC(\rho)\le\lambda_{\max}(S)\,\nu(\rho)\,C(\sigma_\rho), \label{eq:sandwich}
\end{equation}
and $\EC(\rho)\le\Cmax$;

(d) if $C$ is continuous, the supremum in Eq.~\eqref{eq:E} is attained by an instrument with at most $d^2+1$ outcomes.
\end{theorem}
\begin{proof}
(c) Unambiguous discrimination with $a_j=\lambda_{\min}(S)$ contributes $\Ct(\lambda_{\min}(S)M)=\lambda_{\min}(S)\,\nu(\rho)\,C(\sigma_\rho)$, which is the lower bound. Since $\sum_n\hat A_n^\dagger\hat A_n=S\le\lambda_{\max}(S)\id$, the operators $\hat A_n/\sqrt{\lambda_{\max}(S)}$ can be completed to an incoherent instrument by operators that map onto a single basis vector. Strong monotonicity of $C$ and homogeneity of $\Ct$ then give
\begin{align}
\sum_n\Ct\big(\hat A_nM\hat A_n^\dagger\big)&\le\lambda_{\max}(S)\,\Ct(M)\nonumber\\
&=\lambda_{\max}(S)\,\nu(\rho)\,C(\sigma_\rho).
\end{align}
Each $\tau_n$ is supported on at most $d$ basis vectors, so $\EC\le\sum_np_n\Cmax=\Cmax$.

(a) If $\rho$ is free, every $\tau_n$ is incoherent and $\EC(\rho)=0$; otherwise $\sigma_\rho$ is coherent and $\EC(\rho)>0$ by Eq.~\eqref{eq:sandwich}. Each function $M\mapsto\sum_n\Ct(\hat A_nM\hat A_n^\dagger)$ is convex and $M$ is linear in $\rho$, so $\EC$ is convex. For (M3), let $\{\hat K_n\}$ be an FO and $\{\hat A_{l|n}\}_l$ a stochastic orthogonalization for each $n$. The operators $\hat A_{l|n}\hat K_n$ are of the form $\Pi_hD_\xi$ of Eq.~\eqref{eq:Ahat}, and by Eqs.~\eqref{eq:decomp} and~\eqref{eq:FO},
\begin{equation}
\sum_{n,l}\hat K_n^\dagger\hat A_{l|n}^\dagger\hat A_{l|n}\hat K_n=\sum_n\hat K_n^\dagger S\hat K_n=S.
\end{equation}
They form a stochastic orthogonalization whose average coherence on $\rho$ is the sum over $n$ of $p_n$ times the average coherence of $\{\hat A_{l|n}\}_l$ on $\rho_n$. The supremum over the $\hat A_{l|n}$ gives $\sum_np_n\EC(\rho_n)\le\EC(\rho)$, and (M3) with (M4) implies (M2).

(b) For $S=\id$ a stochastic orthogonalization is an incoherent instrument, so $\EC\le C$ by strong monotonicity, and the identity attains $C$.

(d) The pairs $(\hat A^\dagger\hat A,\Ct(\hat AM\hat A^\dagger))\in\mathbb R^{d^2+1}$ scale jointly when $\hat A$ is multiplied by a positive number, so by Carath\'eodory's theorem every sum over outcomes in Eq.~\eqref{eq:E} equals one over at most $d^2+1$ outcomes with the same $\sum_n\hat A_n^\dagger\hat A_n$. After a relabeling of the basis, the maps $h_n$ take values in $d$ fixed labels; by Eq.~\eqref{eq:norm} the outcomes obey $|\xi_{n,j}|\le1$ and form a compact set, on which the average coherence is continuous.
\end{proof}

The measure $\EC$ inherits the meaning of $C$. If $C$ is the relative entropy of coherence, the optimal rate at which incoherent operations distill maximally coherent states~\cite{Winter2016}, then $E_{C_{\rm r}}(\rho)$ is the optimal distillation rate from copies of $\rho$ when each copy is first converted individually by a stochastic orthogonalization. The bound $\EC\le\Cmax$ contrasts with $\Ml$ and the robustness of superposition, which are unbounded as $\lambda_{\min}(S)\to0$~\cite{Torun2026Gram}.

\begin{theorem}[Maximal degree of superposition]\label{thm:max}
Let $C$ be continuous and attain $\Cmax$ only on maximally coherent states (for example $C_{\ell_1}$ or $C_{\rm r}$). For a state $\rho$ the following are equivalent:

(i) $\EC(\rho)=\Cmax$;

(ii) a stochastic orthogonalization converts $\rho$ with certainty into a maximally coherent state;

(iii) $\rho=\ketbra\psi\psi$ is pure, and its coefficient vector $x$ satisfies, for all $j$,
\begin{equation}
\bar x_j\braket{c_j|\psi}=\frac1d,\quad S\ge D_a,\quad a_j=\frac{1}{d|x_j|^2}. \label{eq:max}
\end{equation}
The optimal conversion is then the unambiguous discrimination with success probabilities $a_j$: it never fails on $\ket\psi$, and its output has the population $1/d$ on each of the $d$ labels. Such a state has $\nu(\psi)=\sum_j1/(da_j)>1$ unless $S=\id$.
\end{theorem}
\begin{proof}
(iii)$\Rightarrow$(ii) Since $\braket{c_j|\psi}=(Sx)_j$, Eq.~\eqref{eq:max} gives
\begin{equation}
(Sx)_j=\frac{1}{d\,\bar x_j}=a_jx_j,\quad\text{that is,}\quad (S-D_a)\,x=0.
\end{equation}
The inconclusive outcome of the discrimination with probabilities $a$ therefore never occurs, and the filter leaves $D_a^{1/2}x$, whose entries have modulus $1/\sqrt d$: a maximally coherent state.

(ii)$\Rightarrow$(i) The average coherence is then $\Cmax$, the largest possible value by Theorem~\ref{thm:measure}(c).

(i)$\Rightarrow$(iii) By Theorem~\ref{thm:measure}(d) an optimal instrument exists. Since every $\tau_n$ is supported on at most $d$ basis vectors, $C(\tau_n)\le\Cmax$, and $\sum_np_nC(\tau_n)=\Cmax$ forces every outcome with $p_n>0$ to leave a maximally coherent state on $d$ basis vectors. Then $h_n$ is injective, all $\xi_{n,j}\ne0$, and $\hat A_n$ is invertible; since $\hat A_nM\hat A_n^\dagger$ has rank one, so has $M$, and $\rho$ is pure. Moreover $|\xi_{n,j}x_j|^2=p_n/d$, and summing over these outcomes, whose probabilities add up to one, gives
\begin{equation}
D_a=\sum_{n:\,p_n>0}\hat A_n^\dagger\hat A_n,\quad a_j|x_j|^2=\frac1d.
\end{equation}
The remainder $S-D_a$, the sum over the outcomes with $p_n=0$, is positive semidefinite and has $x^\dagger(S-D_a)x=0$; hence $(S-D_a)x=0$, which is Eq.~\eqref{eq:max}. Finally $a_j\le S_{jj}=1$, and $a_j=1$ for all $j$ requires $S\ge\id$, hence $S=\id$.
\end{proof}

We call a state that satisfies the equivalent conditions of Theorem~\ref{thm:max} a maximal-degree state. By (iii) this notion does not depend on the coherence measure, and by (ii) such a state has $\EC=\Cmax$ for every $C$; for $C$ as in Theorem~\ref{thm:max}, no other state does. In a maximal-degree state the Chirgwin--Coulson weights $\bar x_j\braket{c_j|\psi}$~\cite{Chirgwin1950,Torun2026Gram}, which sum to one, are uniform, and the corresponding discrimination is attainable. With $X=\diag(x)$, Eq.~\eqref{eq:max} is equivalent to $X^\dagger SX\ge\id/d$, that is, to $d\,\lambda_{\min}(X^\dagger SX)=1$, because the entries of $X^\dagger SX$ sum to one. The left-hand side is the optimal probability of obtaining a maximally coherent state from $\ket\psi$~\cite{Torun2026Distill}, so (ii)$\Leftrightarrow$(iii) is the case in which this probability is one. Physically, $\nu(\psi)>1$ means that the components interfere destructively.

\begin{theorem}[Variational characterization and existence]\label{thm:exist}
(a) For every unambiguous discrimination, that is, every $a$ with $a_j\ge0$ and $S\ge D_a$, and every pure state with coefficient vector $x$,
\begin{equation}
d^d\prod_{j}a_j|x_j|^2\le1, \label{eq:amgm}
\end{equation}
with equality iff $\ket\psi$ is a maximal-degree state that this discrimination converts with certainty. Hence all maximal-degree states are converted by the same discrimination, whose success probabilities $a^{\rm opt}$ maximize $\prod_ja_j$, and, if maximal-degree states exist, they are exactly the states that maximize $\prod_j|x_j|^2$.

(b) For $d\le3$, and for $d\le5$ if all overlaps are real, every set of reference states has a maximal-degree state.

(c) For every $d\ge4$ there are reference states without a maximal-degree state.
\end{theorem}
\begin{proof}
(a) By the inequality of arithmetic and geometric means and by $S\ge D_a$,
\begin{align}
\Big(\prod_jd\,a_j|x_j|^2\Big)^{1/d}&\le\sum_ja_j|x_j|^2=x^\dagger D_ax\nonumber\\
&\le x^\dagger Sx=1.
\end{align}
Equality holds iff $a_j|x_j|^2=1/d$ for all $j$ and $(S-D_a)x=0$, because $S-D_a\ge0$. Since $(S-D_a)x=0$ means $\braket{c_j|\psi}=a_jx_j$, these two conditions are Eq.~\eqref{eq:max}, and by Theorem~\ref{thm:max} the discrimination then converts $\ket\psi$ with certainty. If $x$ is the coefficient vector of a maximal-degree state with probabilities $a$, so that $d^d\prod_ja_j|x_j|^2=1$, Eq.~\eqref{eq:amgm} gives $\prod_ja'_j\le\prod_ja_j$ for every attainable $a'$ and $\prod_j|x'_j|^2\le\prod_j|x_j|^2$ for every coefficient vector $x'$, with equality iff $x'$ belongs to a maximal-degree state with the same $a$. The maximizer $a^{\rm opt}$ of $\prod_ja_j$ is unique, because $\sum_j\ln a_j$ is strictly concave on the compact convex set of $a$ with $a_j\ge0$ and $S\ge D_a$.

(b) The set of attainable $a$ contains $\lambda_{\min}(S)(1,\dots,1)/2$ in its interior, so $a^{\rm opt}$ is characterized by the Karush--Kuhn--Tucker conditions~\cite{BoydVandenberghe2004}: $a=a^{\rm opt}$ iff $S\ge D_a$ and some $Z\ge0$ satisfies
\begin{equation}
Z\,(S-D_a)=0,\quad Z_{jj}=1/a_j\quad\text{for all }j. \label{eq:kkt}
\end{equation}
A solution of rank one, $Z=d\,xx^\dagger$, gives $(S-D_{a^{\rm opt}})x=0$ and $a^{\rm opt}_j|x_j|^2=1/d$, that is, a maximal-degree state. The solutions are $Z=UWU^\dagger$, where the columns of $U$ span the kernel of $S-D_{a^{\rm opt}}$ and $W\ge0$ obeys the $d$ real linear equations $(UWU^\dagger)_{jj}=1/a^{\rm opt}_j$. This set is nonempty and compact, so it has an extreme point $W=FF^\dagger$, with $F$ of full column rank $r$. If $r^2>d$, some Hermitian $\Delta\ne0$ of size $r$ satisfies the $d$ equations $(UF\Delta F^\dagger U^\dagger)_{jj}=0$; scaled so that $\id\pm\Delta\ge0$, it gives two further solutions $F(\id\pm\Delta)F^\dagger$ with mean $W$, which contradicts extremality~\cite{Pataki1998,HuangPalomar2010}. Hence $r^2\le d$, and $r=1$ for $d\le3$. For real $S$, $U$, $F$, and $\Delta$ can be taken real, so $r(r+1)/2\le d$ and $r\le2$ for $d\le5$. For $r=2$, the vector $UF\left(\begin{smallmatrix}1\\ i\end{smallmatrix}\right)$ has the squared moduli $(UWU^\dagger)_{jj}$ of its entries, so its outer product with itself is a solution of rank one.

(c) Let $Q$ be the projector onto the span of $u=(1,2,2,i)$ and $v=(2i,1,-1,2)$, and
\begin{equation}
S=\id+\epsilon\,(\id-2Q),\quad 0<\epsilon<1. \label{eq:cex}
\end{equation}
Since $u^\dagger v=0$, $u^\dagger u=v^\dagger v=10$, and $|u_j|^2+|v_j|^2=5$, we have $Q_{jj}=1/2$, and $S$ is a Gram matrix with the eigenvalue $1-\epsilon$ on the span of $u$ and $v$ and $1+\epsilon$ on its complement. $Z=2Q/(1-\epsilon)$ satisfies Eq.~\eqref{eq:kkt} with $a=(1-\epsilon)(1,\dots,1)$, so $a^{\rm opt}=a$, and by (a) a maximal-degree state would have $x=z_1u+z_2v$ with $|x_j|$ independent of $j$. Equal moduli of $x_1=2z_1+z_2$ and $x_2=2z_1-z_2$ require $\mathrm{Re}(\bar z_1z_2)=0$, those of $x_0=z_1+2iz_2$ and $x_3=iz_1+2z_2$ require $\mathrm{Im}(\bar z_1z_2)=0$, and then $|x_0|=|x_1|$ requires $|z_1|=|z_2|$, so $z_1=z_2=0$. For $d>4$, further reference states orthogonal to all others extend the example.
\end{proof}

The product $\prod_j|x_j|^2$ is large when the coefficients are large and evenly spread, that is, when the components interfere destructively and contribute alike, and maximal-degree states exist iff the largest values of the two products saturate Eq.~\eqref{eq:amgm}, $d^d\max_a\prod_ja_j\,\max_x\prod_j|x_j|^2=1$. For the reference states of Eq.~\eqref{eq:cex} a gap remains; for $d=4$ and $d=5$, such a gap requires overlaps that no rephasing of the reference states makes real, that is, a Bargmann invariant~\cite{Bargmann1964,SamuelBhandari1988,MukundaSimon1993}, a cyclic product of overlaps such as $S_{01}S_{12}S_{20}$, that is not real. The coefficient vector of a golden state is an eigenvector of $S$ with the eigenvalue $\lambda_{\min}(S)$ and has entries of equal modulus~\cite{Senyasa2022,Torun2026Gram}; hence a golden state is a balanced maximal-degree state, and its existence forces $a^{\rm opt}=\lambda_{\min}(S)(1,\dots,1)$. For example, for $d=3$, reference states with equal overlap $s=1/2$ have no golden state~\cite{Theurer2017,Torun2026Gram}, whereas their Fourier states $(\ket{c_0}+\omega^{\pm1}\ket{c_1}+\omega^{\mp1}\ket{c_2})/\sqrt{3(1-s)}$, $\omega=e^{2\pi i/3}$, are maximal-degree states, in agreement with Ref.~\cite{Torun2026Distill}.

\section{Two reference states}\label{sec:two}
In this section $d=2$, and we write $S_{01}=s$ with $0<|s|<1$; a state that superposes two states of a larger reference set is treated in Sec.~\ref{sec:general}. A map $h$ on two indices is injective or constant, and the optimization can be carried out for every coherence measure.

\begin{figure*}[tb]
\includegraphics[width=\textwidth]{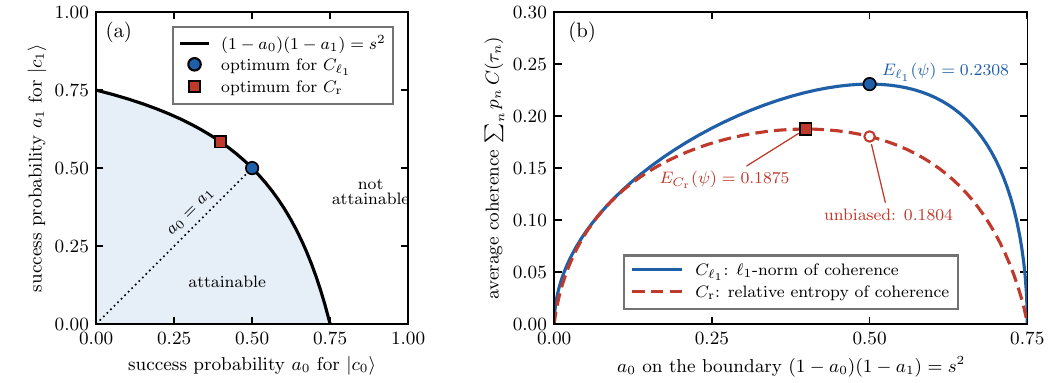}
\caption{Optimal conversion of superposition into coherence for two reference states, $d=2$ (Theorem~\ref{thm:qubit}), for the overlap $s=1/2$ and the state $\ket\psi=(3\ket{c_0}+\ket{c_1})/\sqrt{13}$, which is $\ket{\psi_3}$ of Sec.~\ref{sec:incomp}. (a) Attainable success probabilities of unambiguous discrimination, $(1-a_0)(1-a_1)\ge s^2$ (shaded); the dotted diagonal marks the unbiased discrimination, and the optimum lies on the boundary. (b) Average coherence $\sum_np_nC(\tau_n)$ along the boundary, as a function of $a_0$, which fixes $a_1$. For the $\ell_1$-norm of coherence the optimum is the unbiased discrimination, $a_0=a_1=1-s$ (circle), with $\El(\psi)=(1-s)\Ml(\psi)=0.2308$. For the relative entropy of coherence it is biased toward the smaller component, $(a_0,a_1)=(0.399,0.584)$ (square), with $E_{C_{\rm r}}(\psi)=0.1875$; the unbiased discrimination gives $0.1804$ (open circle).}
\label{fig:one}
\end{figure*}

\begin{theorem}[Two reference states]\label{thm:qubit}
Let $d=2$, with overlap $s$. For every coherence measure $C$ and every state $\rho$ with coefficient matrix $M$,
\begin{equation}
\EC(\rho)=\sup_{a}\,\Ct\big(D_a^{1/2}MD_a^{1/2}\big), \label{eq:qubit}
\end{equation}
where the supremum runs over $a_0,a_1\in[0,1]$ with ${(1-a_0)(1-a_1)=|s|^2}$. If $C$ is continuous, the supremum is attained, and the optimal stochastic orthogonalization is an unambiguous discrimination. For the $\ell_1$-norm of coherence,
\begin{equation}
\El(\rho)=(1-|s|)\,\Ml(\rho), \label{eq:l1}
\end{equation}
attained by the unbiased discrimination, $a_0=a_1=1-|s|$; the robustness of coherence gives the same value.
\end{theorem}
\begin{proof}
If $h_n$ is constant, $\tau_n$ is incoherent and contributes nothing. If it is injective, $\hat A_n$ is a diagonal matrix $D_{\xi_n}$ up to a relabeling of the basis. By Eq.~\eqref{eq:decomp}, the diagonal matrix $D_a=\sum_nD_{\xi_n}^\dagger D_{\xi_n}$ of the injective outcomes obeys $S-D_a\ge0$, and every such $a$ is admissible, because the remainder is a sum of constant-map terms. For given $a$, $D_{\xi_n}=L_nD_a^{1/2}$ with diagonal $L_n$ that form an incoherent instrument, so that
\begin{equation}
\sum_n\Ct\big(D_{\xi_n}MD_{\xi_n}^\dagger\big)\le\Ct\big(D_a^{1/2}MD_a^{1/2}\big),
\end{equation}
with equality for the single filter $D_a^{1/2}$. A diagonal contraction is an incoherent Kraus operator, so this function does not decrease when $a_0$ or $a_1$ increases, and the supremum can be taken on the boundary ${(1-a_0)(1-a_1)=|s|^2}$. For $C_{\ell_1}$ and for the robustness of coherence, $\Ct(Y)=2|Y_{01}|$ on $2\times2$ matrices~\cite{Napoli2016}, so the function is $2\sqrt{a_0a_1}\,|M_{01}|$, and on the boundary
\begin{equation}
\sqrt{a_0a_1}\le1-\sqrt{(1-a_0)(1-a_1)}=1-|s|,
\end{equation}
with equality iff $a_0=a_1$.
\end{proof}

For the relative entropy of coherence (logarithms to base two) and unequal amplitudes, the optimal discrimination is not unbiased but biased toward the component with the smaller amplitude, which makes the outcome more balanced at the price of a lower success probability (Fig.~\ref{fig:one}).

\section{Incomparable superpositions}\label{sec:incomp}

Every degree of superposition constrains the transformations between states. If an FO converts $\rho$ with probability $p$ into a state $\rho'$ that is not free, that is, if its outcomes that leave $\rho'$ have total probability $p$, then (M3) gives $p\,\EC(\rho')\le\EC(\rho)$, because the other outcomes contribute nonnegative terms. Hence the largest such probability obeys
\begin{equation}
p_{\max}(\rho\to\rho')\le\inf_C\frac{\EC(\rho)}{\EC(\rho')}, \label{eq:pmax}
\end{equation}
and a deterministic conversion, one with $p=1$, requires $\EC(\rho)\ge\EC(\rho')$ for every coherence measure $C$.

Two states are incomparable if neither can be converted into the other deterministically by an FO. Pure entangled states of two qubits are all comparable, and incomparable ones exist from two qutrits on~\cite{Nielsen1999}; a catalyst can enable some of these conversions~\cite{JonathanPlenio1999,Datta2023}. Coherence theory behaves alike: conversions of pure states under incoherent operations are governed by majorization of the squared moduli of their coefficients, so all pure states of a qubit are comparable, incomparable ones exist from $d=3$ on, and catalysts again help~\cite{Du2015,*Du2015E,Streltsov2017,Bu2016}. In RTS, incomparable states exist already for $d=2$~\cite{Torun2021}; for example, the even cat state $\propto\ket\alpha+\ket{-\alpha}$ and the Yurke--Stoler state $\propto\ket\alpha+i\ket{-\alpha}$ cannot be converted into each other deterministically~\cite{Torun2026Distill}. Whether a pure state $\ket\psi$ can be converted deterministically into a pure state $\ket{\psi'}$ with nonzero coefficients is decided by Gram majorization~\cite{Torun2026Gram}: the Gram matrix of the components $x_j\ket{c_j}$ of $\ket\psi$ must dominate, in the positive semidefinite order, a mixture of copies of the Gram matrix of the components of $\ket{\psi'}$ with rows and columns permuted alike; for real coefficients and overlaps, conditions of this kind were derived in Ref.~\cite{Torun2021}. The degrees give necessary conditions.

For $d=2$ and a pure state, Eq.~\eqref{eq:l1} gives $\El=2(1-|s|)\,|x_0x_1|$, and Eq.~\eqref{eq:qubit} for the relative entropy of coherence gives
\begin{equation}
E_{C_{\rm r}}=\max_a\,\big(a_0|x_0|^2+a_1|x_1|^2\big)\,H_2\Big(\frac{a_0|x_0|^2}{a_0|x_0|^2+a_1|x_1|^2}\Big), \label{eq:Cr}
\end{equation}
with $a$ on the boundary ${(1-a_0)(1-a_1)=|s|^2}$ and the binary entropy $H_2$. Both are $|x_1|^2$ times a function of $|x_0/x_1|$ and $|s|$, so their ratio $E_{C_{\rm r}}/\El$ does not depend on the relative phase of the coefficients. For $s=1/2$ it decreases from $1$ at $|x_0/x_1|=1$ to $0.92$, $0.81$, and $0.73$ at $|x_0/x_1|=2$, $3$, and $4$, so states with different amplitude ratios can be ordered oppositely by the two degrees. This cannot happen in coherence theory, where all measures order the pure states of a qubit alike. Deterministic conversions obey further constraints, which follow directly from Eq.~\eqref{eq:FO}.

\begin{table}[t]
\caption{Degrees of superposition of the states of Eqs.~\eqref{eq:four} and~\eqref{eq:psi5} for $s=1/2$: $\El$ from Eq.~\eqref{eq:l1}, exact and to three decimals, and $E_{C_{\rm r}}$ from Eq.~\eqref{eq:Cr}, to three decimals.}
\label{tab:incomp}
\begin{ruledtabular}
\begin{tabular}{lccccc}
 & $\psi_1$ & $\psi_2$ & $\psi_3$ & $\psi_4$ & $\psi_5$\\ \hline
$\El$ (exact) & $3/7$ & $4/13$ & $3/13$ & $4/21$ & $2/7$\\
$\El$ & $0.429$ & $0.308$ & $0.231$ & $0.190$ & $0.286$\\
$E_{C_{\rm r}}$ & $0.348$ & $0.223$ & $0.188$ & $0.138$ & $0.262$
\end{tabular}
\end{ruledtabular}
\end{table}

\begin{proposition}[Interference and rigidity]\label{prop:rigidd}
Let an FO convert a pure state $\ket\psi$ deterministically into a pure state $\ket{\psi'}$ whose coefficients are all nonzero.

(a) For every $S$, $\nu(\psi')\le\nu(\psi)$: the interference term $1-\nu(\psi)=\sum_{j\ne k}\bar x_jS_{jk}x_k$ of $\braket{\psi|\psi}$ cannot decrease.

(b) If $S_{jk}=s>0$ for all $j\ne k$ and $\ket\psi$ is in-phase, then $\ket{\psi'}$ is, up to a phase, $\ket\psi$ with its coefficients permuted.
\end{proposition}
\begin{proof}
(a) Each outcome $\hat K_n=\Pi_{h_n}D_{\xi_n}$ with $p_n>0$ leaves $\ket{\psi'}$, $\hat K_nx=\sqrt{p_n}\,e^{i\theta_n}x'$, and these $p_n$ sum to one; since all $x'_j\ne0$, $h_n$ is a bijection, and $|\xi_{n,j}x_j|^2=p_n|x'_{h_n(j)}|^2$. The outcomes with $p_n=0$ have $\hat K_nx=0$. Multiplying $\sum_n|\xi_{n,j}|^2=1$, the diagonal of Eq.~\eqref{eq:FO}, by $|x_j|^2$ gives
\begin{equation}
|x_j|^2=\sum_{n:\,p_n>0}p_n\,|x'_{h_n(j)}|^2+\kappa_j, \label{eq:diag}
\end{equation}
where $\kappa_j\ge0$ is the sum of $|\xi_{n,j}x_j|^2$ over the outcomes with $p_n=0$; summing over $j$ gives $\nu(\psi)=\nu(\psi')+\sum_j\kappa_j$.

(b) For in-phase $\ket\psi$, $1-\nu(\psi)=s\,\Ml(\psi)$, whereas $1-\nu(\psi')\le s\,\Ml(\psi')$, with equality iff $\ket{\psi'}$ is in-phase, and $\Ml(\psi')\le\Ml(\psi)$, since $\El=(1-s)\Ml$ (Corollary~\ref{cor:sym}) cannot increase. By (a), these inequalities are equalities: $\ket{\psi'}$ is in-phase, all $\kappa_j$ vanish, and $\sum_j|x_j|=\sum_j|x'_j|$, since $\Ml=(\sum_j|x_j|)^2-\nu$ for pure states. By Eq.~\eqref{eq:diag} and the concavity of the square root, $|x_j|\ge\sum_np_n|x'_{h_n(j)}|$, and the right-hand sides sum to $\sum_j|x'_j|$. Hence equality holds for every $j$, so that $|x'_{h_n(j)}|=|x_j|$ whenever $p_n>0$; since both states are in-phase, $x'$ is $x$ with its entries permuted by $h_n$, up to a phase.
\end{proof}

\begin{figure}[t]
\includegraphics[width=\columnwidth]{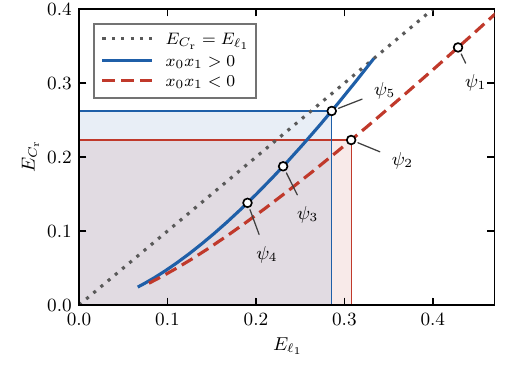}
\caption{Degrees of superposition of the pure states $x_0\ket{c_0}+x_1\ket{c_1}$ for $s=1/2$, from Theorem~\ref{thm:qubit}. Each state is a point $(\El,E_{C_{\rm r}})$. Solid blue: $x_0x_1>0$; dashed red: $x_0x_1<0$; along each curve $|x_0/x_1|$ increases from $1$ to $14$ toward the lower left, and the dashed curve starts outside the frame, at $(1,1)$. Dotted: $E_{C_{\rm r}}=\El$. Circles: the states of Eqs.~\eqref{eq:four} and~\eqref{eq:psi5}. Since neither degree can increase, a deterministic conversion of a state requires the target in the quadrant to its lower left, shaded for $\psi_2$ (red) and $\psi_5$ (blue). Neither of these two states lies in the quadrant of the other, so the degrees alone show that they are incomparable.}
\label{fig:incomp}
\end{figure}

Since $\nu(\psi)>1$ means that the components interfere destructively (Sec.~\ref{sec:measure}), part (a) is the principle of Ref.~\cite{Torun2026Distill} that FO cannot make the components interfere more destructively with certainty while keeping all of them; there it fixes the optimal conversion probability for every source symmetric under permutations of its components. For equal positive overlaps the permutations of the reference states are unitary FO, so two in-phase states with nonzero coefficients are incomparable unless they are related by a permutation: constructively interfering superpositions are rigid and form continuous families of pairwise incomparable states. For $d=2$, a rephasing of $\ket{c_1}$ makes the overlap positive, and the deterministic case takes an explicit form.

\begin{proposition}[Deterministic conversions for $d=2$]\label{prop:rigid}
Let $d=2$ with overlap $s>0$, and let $\ket\psi$ and $\ket{\psi'}$ be pure states with coefficient vectors $x$ and $x'$, all coefficients nonzero. If an FO converts $\ket\psi$ deterministically into $\ket{\psi'}$, then
\begin{equation}
\begin{gathered}
\mathrm{Re}(\bar x'_0x'_1)\ge\mathrm{Re}(\bar x_0x_1),\\
\big||x_0|^2-|x_1|^2\big|\le\big||x'_0|^2-|x'_1|^2\big|.
\end{gathered} \label{eq:conv}
\end{equation}
For real coefficients these two conditions are also sufficient.
\end{proposition}
\begin{proof}
The bijections of two indices are the identity and the exchange; let $w$ be the total probability of the outcomes with the identity. An outcome with $p_n=0$ has $\xi_{n,0}x_0=-\xi_{n,1}x_1$ if $h_n$ is constant and $\xi_{n,j}x_j=0$ otherwise, so $\kappa_0=\kappa_1=:\kappa$ in Eq.~\eqref{eq:diag}. This and the off-diagonal entry of Eq.~\eqref{eq:FO}, multiplied by $\bar x_0x_1$, give
\begin{equation}
\begin{aligned}
|x_j|^2&=w\,|x'_j|^2+(1-w)\,|x'_{1-j}|^2+\kappa,\quad j=0,1,\\
s\,\bar x_0x_1&=s\,\big[w\,\bar x'_0x'_1+(1-w)\,\bar x'_1x'_0\big]-\kappa.
\end{aligned} \label{eq:three}
\end{equation}
The real part of the last equation gives the first condition in Eq.~\eqref{eq:conv}, and the difference of the first two, $|x_0|^2-|x_1|^2=(2w-1)(|x'_0|^2-|x'_1|^2)$, gives the second. Conversely, let the coefficients be real and satisfy Eq.~\eqref{eq:conv}. Then $\kappa=s(x'_0x'_1-x_0x_1)\ge0$ solves the last equation, the sum of the first two holds by the normalization $x_0^2+x_1^2+2s\,x_0x_1=1$ of both states, and their difference holds for some $w\in[0,1]$. By Eq.~\eqref{eq:three}, the three outcomes with $\xi_j=\sqrt w\,x'_j/x_j$ and the identity, $\xi_j=\sqrt{1-w}\,x'_{1-j}/x_j$ and the exchange, and $\xi=\sqrt\kappa\,(1/x_0,-1/x_1)$ and a constant map then satisfy Eq.~\eqref{eq:FO}, and they convert $\ket\psi$ into $\ket{\psi'}$.
\end{proof}

A deterministic conversion thus can lower neither the interference term $2s\,\mathrm{Re}(\bar x_0x_1)$ of $\braket{\psi|\psi}$ nor the imbalance $\big||x_0|^2-|x_1|^2\big|$ of the two weights, whereas $\El$ forbids raising $|x_0x_1|$. For an anti-phase state, $\bar x_0x_1<0$, the first constraint and $\El$ point the same way, and conversions that lower $|x_0x_1|$ can be possible; for an in-phase state they fix $\bar x_0x_1$, in accordance with Proposition~\ref{prop:rigidd}(b). Proposition~\ref{prop:rigid} agrees with Gram majorization~\cite{Torun2026Gram}. For real coefficients, the Gram matrices $G$ and $G'$ of the components, $G_{jk}=x_jS_{jk}x_k$ and $G'_{jk}=x'_jS_{jk}x'_k$, have entries that sum to one, so any positive semidefinite remainder $G-wG'-(1-w)G'_{\rm ex}$, with $G'_{\rm ex}$ obtained from $G'$ by exchanging the two indices, has vanishing row sums and equals $\kappa\left(\begin{smallmatrix}1&-1\\-1&1\end{smallmatrix}\right)$; its entries are Eq.~\eqref{eq:three}, and Gram majorization reduces to Eq.~\eqref{eq:conv}. Reference~\cite{Torun2026Distill} gives the optimal probability of every conversion for $d=2$, and Proposition~\ref{prop:rigid} states its deterministic case explicitly. For $s=0$, $\kappa=0$ and only the second condition remains, the majorization condition of coherence theory~\cite{Du2015}.

We take $s=1/2$ and the four states of Ref.~\cite{Torun2021}, with $\ket{c_0},\ket{c_1}$ in place of $\ket{c_1},\ket{c_2}$ there,
\begin{equation}
\begin{aligned}
\ket{\psi_1}&=\frac{3\ket{c_0}-\ket{c_1}}{\sqrt7}, &\quad \ket{\psi_2}&=\frac{4\ket{c_0}-\ket{c_1}}{\sqrt{13}},\\
\ket{\psi_3}&=\frac{3\ket{c_0}+\ket{c_1}}{\sqrt{13}}, &\quad \ket{\psi_4}&=\frac{4\ket{c_0}+\ket{c_1}}{\sqrt{21}},
\end{aligned} \label{eq:four}
\end{equation}
together with the state
\begin{equation}
\ket{\psi_5}=\frac{2\ket{c_0}+\ket{c_1}}{\sqrt7}. \label{eq:psi5}
\end{equation}
Table~\ref{tab:incomp} and Fig.~\ref{fig:incomp} give their degrees. Both degrees order $\psi_1,\psi_2,\psi_3,\psi_4$ as listed and therefore exclude every conversion into a state earlier in this order. Propositions~\ref{prop:rigidd} and~\ref{prop:rigid} decide all twenty conversions from one of the five states into another. The in-phase states $\psi_3$, $\psi_4$, and $\psi_5$ are rigid, so none of them can be converted into another of the five states. In particular, $\psi_3\to\psi_4$ is impossible~\cite{Torun2021} although both degrees, and the $\ell_1$-measure of superposition, are larger for $\psi_3$: the conversion would have to reduce the constructive interference of $\psi_3$. Of the conversions from the anti-phase states $\psi_1$ and $\psi_2$, those into $\psi_3$, $\psi_4$, and $\psi_5$ would lower the imbalance, which is $8/7$ and $15/13$ for $\psi_1$ and $\psi_2$ and at most $5/7$ for the targets, and $\psi_2\to\psi_1$ would lower the interference term. Only $\psi_1\to\psi_2$ is possible deterministically, with $w=209/210$ and $\kappa=11/182$ in Eq.~\eqref{eq:three}, as realized explicitly in Ref.~\cite{Torun2021}. Hence $\psi_1$ and $\psi_2$ are comparable, although the conversion cannot be reversed, and every other pair of the five states is incomparable: neither state can be converted into the other with certainty. For each of these nine pairs, the degrees exclude one direction, and rigidity or the imbalance excludes the other; only for $\psi_2,\psi_5$ do the degrees exclude both directions.

The pair $\psi_2,\psi_5$ shows what different coherence measures add. The two degrees order it oppositely: $\El(\psi_5)=2/7<\El(\psi_2)=4/13$, but $E_{C_{\rm r}}(\psi_5)=0.262>E_{C_{\rm r}}(\psi_2)=0.223$. Hence $\El$ excludes $\psi_5\to\psi_2$, $E_{C_{\rm r}}$ excludes $\psi_2\to\psi_5$, and the degrees alone show that $\psi_2$ and $\psi_5$ are incomparable, whereas the $\ell_1$-measure of superposition, proportional to $\El$ for $d=2$, excludes only the first conversion. Equation~\eqref{eq:pmax} bounds both, $p_{\max}(\psi_2\to\psi_5)\le0.850$ and $p_{\max}(\psi_5\to\psi_2)\le13/14$.

\section{Any number of reference states}\label{sec:general}
For $d\ge3$ an outcome can also discriminate the reference states partially. We first bound $\El$ by a sum over pairs of components and then determine when the bound is attained.

\begin{theorem}[Pairwise bound]\label{thm:bounds}
For every set of reference states and every state $\rho$,
\begin{equation}
\lambda_{\min}(S)\,\Ml(\rho)\le\El(\rho)\le\sum_{j\ne k}\big(1-|S_{jk}|\big)\,|M_{jk}|. \label{eq:bounds}
\end{equation}
\end{theorem}
\begin{proof}
The lower bound is Eq.~\eqref{eq:sandwich} for $C=C_{\ell_1}$. The off-diagonal element $(\mu,\mu')$ of $\hat A_nM\hat A_n^\dagger$ is the sum of $\xi_{n,j}M_{jk}\bar\xi_{n,k}$ over $h_n(j)=\mu$, $h_n(k)=\mu'$, so that
\begin{equation}
\begin{aligned}
\sum_n\Ct_{\ell_1}\big(\hat A_nM\hat A_n^\dagger\big)\le\sum_{j\ne k}&|M_{jk}|\\
\times&\sum_{n:\,h_n(j)\ne h_n(k)}|\xi_{n,j}|\,|\xi_{n,k}|.
\end{aligned} \label{eq:pair}
\end{equation}
Fix $j\ne k$ and let $g_j$ and $g_k$ be the sums of $|\xi_{n,j}|^2$ and $|\xi_{n,k}|^2$ over the outcomes with $h_n(j)=h_n(k)$, which group the two states together. By Eq.~\eqref{eq:decomp} and the Cauchy--Schwarz inequality, $|S_{jk}|\le\sqrt{g_jg_k}$. By Eq.~\eqref{eq:norm}, the sums of $|\xi_{n,j}|^2$ and $|\xi_{n,k}|^2$ over the remaining outcomes are $1-g_j$ and $1-g_k$, and the Cauchy--Schwarz inequality bounds the inner sum in Eq.~\eqref{eq:pair} by
\begin{equation}
\sqrt{(1-g_j)(1-g_k)}\le1-\sqrt{g_jg_k}\le1-|S_{jk}|, \label{eq:chain}
\end{equation}
which gives the upper bound in Eq.~\eqref{eq:bounds}.
\end{proof}

We call the right-hand side of Eq.~\eqref{eq:bounds} the pairwise bound. Its weight $1-|S_{jk}|$ is the optimal probability of unambiguously discriminating $\ket{c_j}$ and $\ket{c_k}$ alone~\cite{Ivanovic1987,Dieks1988,Peres1988}, which also measures path distinguishability in multipath interference~\cite{Bera2015}: the interference term between two components is converted into coherence with at most the probability with which its two reference states can be told apart. For $d\ge3$ the pairs cannot in general be discriminated optimally at once; by the equality conditions in the proof, the bound is attained for a pair with $M_{jk}\ne0$ only if every outcome treats $\ket{c_j}$ and $\ket{c_k}$ alike, $|\xi_{n,j}|=|\xi_{n,k}|$, and $\bar\xi_{n,j}\xi_{n,k}$ has the same phase in all outcomes that group the two states together (Theorem~\ref{thm:grouping} below). The two bounds determine $\El$ whenever they coincide.

\begin{corollary}[Uniform overlaps]\label{cor:sym}
The two bounds in Eq.~\eqref{eq:bounds} coincide for every state iff
\begin{equation}
S=(1-s)\,\id+s\,\zeta\zeta^\dagger,\quad |\zeta_j|=1, \label{eq:sym}
\end{equation}
for some $0\le s<1$, that is, iff all overlaps have the modulus $s$ and become equal to $s$ after a rephasing of the reference states; we then say that the reference states have uniform overlaps. Then, for every state $\rho$,
\begin{equation}
\El(\rho)=(1-s)\,\Ml(\rho), \label{eq:symE}
\end{equation}
attained by the unbiased discrimination, $a_j=1-s$.
\end{corollary}
\begin{proof}
For $S$ of the form~\eqref{eq:sym}, $\lambda_{\min}(S)=1-s=1-|S_{jk}|$ for all $j\ne k$, so the bounds coincide. Conversely, states whose coefficient matrix has a single pair of nonzero off-diagonal entries $M_{jk}=\bar M_{kj}$ show that coincidence requires $1-|S_{jk}|=\lambda_{\min}(S)$ for every pair, hence $|S_{jk}|=s$. For $s=0$ this gives $S=\id$; for $s>0$, $\Gamma=[S-(1-s)\id]/s$ is positive semidefinite with unit diagonal and unimodular entries; it is the Gram matrix of unit vectors with unimodular overlaps, which are therefore parallel, so $\Gamma=\zeta\zeta^\dagger$ with $|\zeta_j|=1$.
\end{proof}

For $d=2$ every Gram matrix has this form, with $|S_{01}|$ in place of $s$, and Eq.~\eqref{eq:symE} is Eq.~\eqref{eq:l1}. For $d\ge3$, Eq.~\eqref{eq:symE} gives $\El$ in closed form for all states, mixed or pure and with arbitrary phases. For example, the maximally mixed state $\id/d$ of $\Hs$, a superposition state (Sec.~\ref{sec:setting}) with $M=S^{-1}/d$, has for uniform overlaps the degree of superposition
\begin{equation}
\El(\id/d)=\frac{(d-1)\,s}{1+(d-1)\,s}, \label{eq:mixedd}
\end{equation}
which grows with the overlap and tends to $(d-1)/d$, the fraction $1/d$ of the largest value $C_{\ell_1,\max}(d)=d-1$, as the reference states approach each other.

\begin{theorem}[Grouping]\label{thm:grouping}

(a) Let $w_P\ge0$ be weights with $\sum_Pw_PB_P\le S$. Then, for every coherence measure $C$ and every state $\rho$,
\begin{equation}
\EC(\rho)\ge\sum_Pw_P\,\Ct\big(\Pi_PM\Pi_P^\dagger\big). \label{eq:grouping}
\end{equation}

(b) If $S=\sum_Pw_PB_P$ with $w_P\ge0$, which implies $S_{jk}\ge0$, then for every in-phase state,
\begin{equation}
\El(\rho)=\sum_{j\ne k}(1-S_{jk})\,M_{jk}, \label{eq:exact}
\end{equation}
attained by the instrument $\{\sqrt{w_P}\,\Pi_P\}$.

(c) Conversely, if the pairwise bound is attained for a state with $M_{jk}\ne0$ for all $j\ne k$, then there are weights $w_P\ge0$ with $\sum_Pw_P=1$ such that
\begin{equation}
|S_{jk}|=\sum_Pw_P\,(B_P)_{jk}\quad\text{for all }j,k. \label{eq:converse}
\end{equation}
\end{theorem}
\begin{proof}
(a) The operator $\sqrt{w_P}\,\Pi_P$ has the form $\Pi_hD_\xi$ of Eq.~\eqref{eq:Ahat} with $\xi_{P,j}=\sqrt{w_P}$, and $w_P\Pi_P^\dagger\Pi_P=w_PB_P$. The positive semidefinite remainder $S-\sum_Pw_PB_P$ is realized by operators with a constant map and contributes no coherence. Equation~\eqref{eq:E} gives Eq.~\eqref{eq:grouping}.

(b) For $M_{jk}\ge0$ the entries of $\Pi_PM\Pi_P^\dagger$ are sums of nonnegative numbers, so the right-hand side of Eq.~\eqref{eq:grouping} for $C=C_{\ell_1}$ is
\begin{equation}
\sum_Pw_P\,\Ct_{\ell_1}\big(\Pi_PM\Pi_P^\dagger\big)=\sum_{j\ne k}M_{jk}\sum_{P:\,j\not\sim_Pk}w_P,
\end{equation}
where $j\sim_Pk$ means that $j$ and $k$ lie in the same class of $P$. The diagonal and the off-diagonal entries of $S=\sum_Pw_PB_P$ give
\begin{equation}
\sum_Pw_P=1,\quad \sum_{P:\,j\sim_Pk}w_P=S_{jk},
\end{equation}
so $\sum_{P:\,j\not\sim_Pk}w_P=1-S_{jk}$, and the lower bound~\eqref{eq:grouping} equals the pairwise bound.

(c) Let an instrument attain the bound [Theorem~\ref{thm:measure}(d)]. Since $M_{jk}\ne0$, every inequality in the proof of Theorem~\ref{thm:bounds} is an equality for every pair $j\ne k$. In Eq.~\eqref{eq:chain}, the first one gives $g_j=g_k$; the Cauchy--Schwarz equality $|S_{jk}|=\sqrt{g_jg_k}$ gives $\xi_{n,k}=e^{i\theta_{jk}}\xi_{n,j}$ on the outcomes that group $j$ with $k$, and the Cauchy--Schwarz equality on the remaining outcomes, on which the sums of $|\xi_{n,j}|^2$ and $|\xi_{n,k}|^2$ are both $1-g_j=1-g_k$, gives $|\xi_{n,j}|=|\xi_{n,k}|$. Hence $|\xi_{n,j}|=|\xi_{n,0}|$ for all $j$, so that $\sum_n|\xi_{n,0}|^2=1$ by Eq.~\eqref{eq:norm}, and
\begin{equation}
|S_{jk}|=g_j=\sum_n|\xi_{n,0}|^2\,(B_{P_n})_{jk},
\end{equation}
where $P_n$ is the partition defined by $h_n$. Collecting outcomes with equal partitions gives the weights $w_P$.
\end{proof}

For nonnegative overlaps and coefficient matrices with positive entries, the pairwise bound is thus attained iff $S$ is a mixture of groupings, $S=\sum_Pw_PB_P$ with $w_P\ge0$. Then, up to an isometry, $\ket{c_j}=\sum_P\sqrt{w_P}\,\ket P\ket{\mu_P(j)}$ with orthonormal $\ket P$ and $\mu_P(j)$ the class of $j$ in $P$: each overlap is the probability that a random grouping puts the two states into the same class, and the optimal orthogonalization reads out the grouping. For every Gram matrix, maximizing the right-hand side of Eq.~\eqref{eq:grouping} over the weights $w_P$ gives a computable lower bound on $\EC$.

\begin{proposition}[Mixtures of groupings]\label{prop:mixture}
Let $S_{jk}\ge0$ for all $j,k$.

(a) If $S=\sum_Pw_PB_P$ with $w_P\ge0$, then
\begin{equation}
S_{jk}+S_{kl}\le1+S_{jl}\quad\text{for all }j,k,l, \label{eq:triangle}
\end{equation}
that is, $1-S_{jk}$ is a distance.

(b) For $d=3$ this condition is also sufficient: $S$ is a mixture of groupings iff the sum of the two largest overlaps is at most $1$ plus the smallest. The weights of the five partitions are
\begin{equation}
\begin{aligned}
w_{\{012\}}&=\gamma,\\
w_{\{01|2\}}&=S_{01}-\gamma,\\
w_{\{02|1\}}&=S_{02}-\gamma,\\
w_{\{0|12\}}&=S_{12}-\gamma,\\
w_{\{0|1|2\}}&=1+2\gamma-S_{01}-S_{02}-S_{12},
\end{aligned} \label{eq:weights3}
\end{equation}
for any $\gamma$ with
\begin{equation}
\max\Big\{0,\frac{S_{01}+S_{02}+S_{12}-1}{2}\Big\}\le\gamma\le\min_{j<k}S_{jk}. \label{eq:gamma}
\end{equation}

(c) For every $d$, hierarchical overlaps, ${S_{jl}\ge\min\{S_{jk},S_{kl}\}}$ for all $j,k,l$, give a mixture,
\begin{equation}
S=\int_0^1B_{P(y)}\,dy, \label{eq:hier}
\end{equation}
where $P(y)$ puts $j$ and $k$ into the same class iff $S_{jk}\ge y$.
\end{proposition}
\begin{proof}
(a) Each $B_P$ satisfies the inequalities, because belonging to the same class is transitive, and mixtures preserve them, since $\sum_Pw_P=1$.

(b) The five partitions of three indices with the stated weights reproduce $S$, and the weights are nonnegative for some $\gamma$ iff $S_{01}+S_{02}+S_{12}-1\le2\min_{j<k}S_{jk}$, which is the stated condition.

(c) For hierarchical overlaps the relation $S_{jk}\ge y$ is transitive for every $y\in(0,1]$, hence a partition, and $\int_0^1[S_{jk}\ge y]\,dy=S_{jk}$, where $[\,\cdot\,]$ is $1$ if the condition holds and $0$ otherwise.
\end{proof}

For $d\ge4$ the triangle inequalities~\eqref{eq:triangle} are not sufficient. The Gram matrix with $S_{0j}=1/2$ for $j=1,2,3$ and $S_{jk}=0$ otherwise is positive definite and satisfies them, but $\sum_{j=1}^3S_{0j}-\sum_{1\le j<k\le3}S_{jk}=3/2$, whereas every $B_P$, and hence every mixture of groupings, gives at most $1$: the first sum is the number of the indices $1,2,3$ in the class of $0$, and the second is at least the number of pairs among them. Adding reference states orthogonal to all others extends the example to every $d>4$.

\begin{figure*}[tb]
\begin{minipage}[t]{0.485\textwidth}
\includegraphics[width=\linewidth]{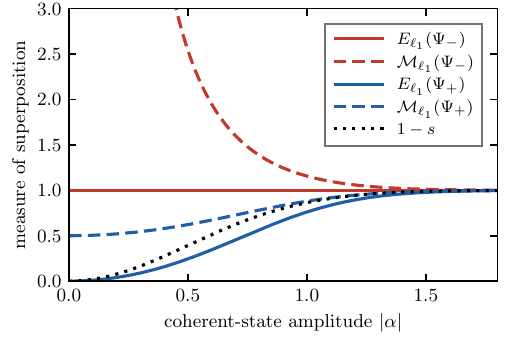}
\caption{Two-component cat states $\ket{\Psi_\pm}\propto\ket\alpha\pm\ket{-\alpha}$, with overlap $s=e^{-2|\alpha|^2}$. Solid: $\El$ of Eq.~\eqref{eq:cat}; dashed: the $\ell_1$-measure of superposition $\Ml(\Psi_\pm)=1/(1\pm s)$; red: odd, blue: even state. Dotted: the factor $1-s$ in Eq.~\eqref{eq:l1}, the optimal probability of unambiguously discriminating $\ket\alpha$ and $\ket{-\alpha}$. The odd cat state is a maximal-degree state for every $|\alpha|$ (Theorem~\ref{thm:max}); as $|\alpha|\to0$ its $\ell_1$-measure diverges, while $\El=1$.}
\label{fig:cat}
\end{minipage}\hfill
\begin{minipage}[t]{0.485\textwidth}
\includegraphics[width=\linewidth]{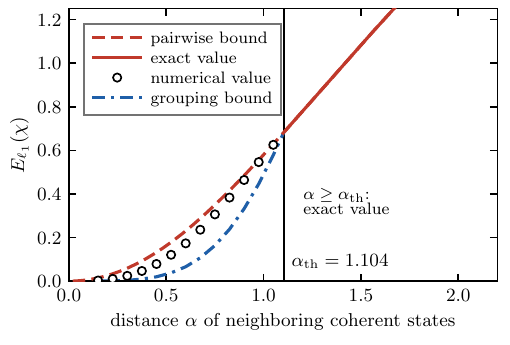}
\caption{Uniform superposition $\chi$ of three coherent states $\ket{-\alpha},\ket0,\ket\alpha$. Solid red: exact value, Eq.~\eqref{eq:line}, which holds above the threshold amplitude $\alpha_{\rm th}=1.104$ (vertical line), where the Gram matrix is a mixture of groupings (Proposition~\ref{prop:mixture}); dashed red: pairwise bound of Theorem~\ref{thm:bounds}; circles: $\El$ obtained by numerical maximization of Eq.~\eqref{eq:E}; dash-dotted blue: grouping bound~\eqref{eq:grouping}, maximized over the weights $w_P$. For $\alpha\ge\alpha_{\rm th}$ both bounds equal the exact value; below $\alpha_{\rm th}$ the numerical values lie between the two bounds, strictly below the pairwise bound [Theorem~\ref{thm:grouping}(c)].}
\label{fig:line}
\end{minipage}
\end{figure*}

The set of mixtures of groupings is the clique-partitioning polytope of combinatorial optimization~\cite{Grotschel1990}; membership is decided by a linear program in the weights $w_P$. For hierarchical overlaps, that is, nested clusters of reference states, Proposition~\ref{prop:mixture}(c) with Theorem~\ref{thm:grouping}(b) gives $\El$ in closed form for all in-phase states and every $d$. If the matrix $|S|$ of the moduli $|S_{jk}|$ is not a mixture of groupings, $\El$ lies strictly below the pairwise bound for every state with $M_{jk}\ne0$ for all $j\ne k$ [Theorem~\ref{thm:grouping}(c)]; for the Gram matrix of the preceding paragraph and the uniform superposition, three outcomes, each grouping $\ket{c_0}$ with one of the other reference states, give $\El\ge(3+10\sqrt6)/28=0.982$, against the pairwise bound $9/7=1.286$. The numerical values in Figs.~\ref{fig:line} and~\ref{fig:circle} were obtained by maximizing Eq.~\eqref{eq:E} over instruments with up to two outcomes for each partition of the index set; a second outcome did not increase them.

A closed form of $\El$ for all Gram matrices is not to be expected: by Theorem~\ref{thm:grouping}, already for in-phase states the question whether the pairwise bound is attained is membership in the clique-partitioning polytope, over which optimization is NP-hard~\cite{Grotschel1990,Bansal2004}.

The degree of a state depends on the whole reference set, not only on the reference states in its expansion. If $\rho$ superposes only the reference states of a subset, restricting every $\hat A_n$ to the corresponding columns gives, by Eq.~\eqref{eq:decomp}, a stochastic orthogonalization of that subset with the same outputs. Hence $\EC(\rho)$ cannot increase when reference states are added, because the free operations must then respect more states. The decrease can be strict. Let $\ket{c_0}$ and $\ket{c_1}$ be orthogonal, let $\ket{c_2}$ have the overlap $S_{02}=S_{12}\in(0,1/\sqrt2)$ with both, and let $\ket\psi=(\ket{c_0}+\ket{c_1})/\sqrt2$, so that $\El(\psi)=1$ without $\ket{c_2}$. With $\ket{c_2}$, $\El(\psi)=1$ iff $S_{02}\le1/2$. For $S_{02}\le1/2$, two outcomes with $\xi=(1/\sqrt2,1/\sqrt2,\sqrt2\,S_{02})$, which group $\ket{c_2}$ with $\ket{c_0}$ and with $\ket{c_1}$, respectively, and a constant map for the rest of $\ket{c_2}$ attain it. Conversely, let $\El(\psi)=1$, attained by an instrument [Theorem~\ref{thm:measure}(d)]. It saturates Eq.~\eqref{eq:bounds}, so by the equality conditions $|\xi_{n,0}|=|\xi_{n,1}|$ in every outcome, and, since $S_{01}=0$, every outcome with $\xi_{n,0}\ne0$ separates $\ket{c_0}$ and $\ket{c_1}$; such an outcome groups $\ket{c_2}$ with at most one of them. Let $g_0$ and $g_1$ be the sums of $|\xi_{n,0}|^2$ over the outcomes with $\xi_{n,0}\ne0$ that group $\ket{c_2}$ with $\ket{c_0}$ and with $\ket{c_1}$, respectively, and $g'_0$ and $g'_1$ the sums of $|\xi_{n,2}|^2$ over the same outcomes. Then $g_0+g_1\le1$ and $g'_0+g'_1\le1$ by Eq.~\eqref{eq:norm}, and Eq.~\eqref{eq:decomp} with the Cauchy--Schwarz inequality gives $S_{02}^2=S_{02}S_{12}\le\sqrt{g_0g'_0g_1g'_1}\le(g_0+g_1)(g'_0+g'_1)/4\le1/4$.

\section{Measures weighted by distinguishability}\label{sec:weighted}
The pairwise bound of Theorem~\ref{thm:bounds} is a superposition measure in closed form, and one of a family.

\begin{theorem}[Weighted measures]\label{thm:weighted}
Let $\varphi$ be a concave, nonincreasing function on $[0,1]$ with $\varphi(1)=0$ that does not vanish identically. Then
\begin{equation}
\mathcal M_\varphi(\rho)=\sum_{j\ne k}\varphi(|S_{jk}|)\,|M_{jk}| \label{eq:weighted}
\end{equation}
is a superposition measure, that is, it satisfies (M1)--(M4).
\end{theorem}
\begin{proof}
Such a function is nonnegative and has $\varphi(0)>0$, since otherwise it would vanish identically; by concavity, $\varphi(t)\ge(1-t)\varphi(0)>0$ on $[0,1)$. Since $|S_{jk}|<1$ for $j\ne k$, $\mathcal M_\varphi$ vanishes iff $M$ is diagonal; it is a seminorm of $M$, hence convex. Let $\{\hat K_n=\Pi_{h_n}D_{\xi_n}\}$ be an FO. As in Eq.~\eqref{eq:pair},
\begin{equation}
\sum_np_n\mathcal M_\varphi(\rho_n)\le\sum_{j\ne k}|M_{jk}|\sum_n\eta_n\,\varphi(\beta_n),
\end{equation}
with $\eta_n=|\xi_{n,j}||\xi_{n,k}|$ and $\beta_n=|S_{h_n(j)h_n(k)}|$ for the pair $j,k$; the terms with $h_n(j)=h_n(k)$ vanish because $\varphi(1)=0$. The diagonal of Eq.~\eqref{eq:FO} gives $\sum_n|\xi_{n,j}|^2=1$, hence, by the Cauchy--Schwarz inequality, $\eta=\sum_n\eta_n\le1$, and its off-diagonal entries give $|S_{jk}|\le\sum_n\eta_n\beta_n$. Since $\varphi$ is concave with $\varphi(0)\ge0$ and nonincreasing,
\begin{align}
\sum_n\eta_n\varphi(\beta_n)&\le\sum_n\eta_n\varphi(\beta_n)+(1-\eta)\varphi(0)\nonumber\\
&\le\varphi\Big(\sum_n\eta_n\beta_n\Big)\le\varphi(|S_{jk}|),
\end{align}
which is (M3); with (M4) it implies (M2).
\end{proof}

In $\mathcal M_\varphi$, an interference term counts only as far as its two reference states can be distinguished: $\varphi(t)=1-t$, written $\mathcal M_{1-t}$, weights it by the optimal probability of unambiguous discrimination and gives the pairwise bound, which equals $\El$ in the cases of Theorems~\ref{thm:qubit} and~\ref{thm:grouping}(b) and Corollary~\ref{cor:sym}; $\varphi(t)=\sqrt{1-t^2}$ weights it by the trace distance, which governs minimum-error discrimination~\cite{Helstrom1976}, as in the relations between coherence and path information~\cite{Englert1996,Bagan2016}. The $\ell_1$-measure of superposition~\cite{Theurer2017} lies outside this family, because its constant weight $\varphi=1$ does not vanish at $t=1$, and, unlike $\El$ and $\mathcal M_{1-t}$, it diverges for the odd cat state as the amplitude tends to zero (Sec.~\ref{sec:examples}).

\section{Examples}\label{sec:examples}
We treat three families of superpositions of coherent states $\ket\alpha$ of a field mode~\cite{Glauber1963}, with $\braket{\alpha|\alpha'}=e^{-|\alpha|^2/2-|\alpha'|^2/2+\bar\alpha\alpha'}$. The coherent states are the reference states and thus free; the resource lies in their superpositions. In this section, $n$ denotes a photon number and $\ket n$ the $n$-photon Fock state.

\begin{figure*}[tb]
\includegraphics[width=\textwidth]{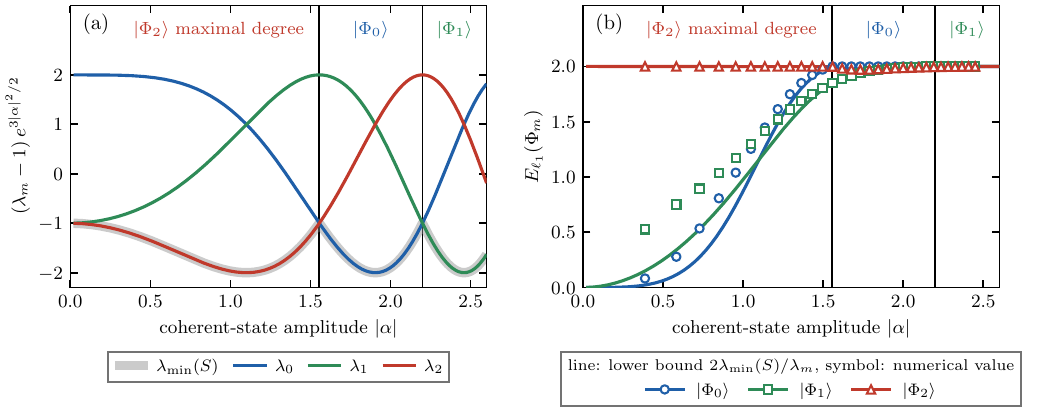}
\caption{Rotation-symmetric cat states $\ket{\Phi_m}$, Eq.~\eqref{eq:circle}, of three coherent states on a circle, $m=0$ (blue), $1$ (green), $2$ (red). (a) Eigenvalues $\lambda_m$ of the Gram matrix, shown as $(\lambda_m-1)e^{3|\alpha|^2/2}$ to make their exponentially small differences visible; the gray band marks the smallest, whose state is of maximal degree: $\Phi_2$ for $|\alpha|<1.555$, $\Phi_0$ for $1.555<|\alpha|<2.199$, and $\Phi_1$ for larger $|\alpha|$ in the range shown (vertical lines). (b) Degree of superposition $\El(\Phi_m)$. Lines: lower bound $2\lambda_{\min}(S)/\lambda_m$, Eq.~\eqref{eq:sandwich}; symbols: numerical maximization of Eq.~\eqref{eq:E}. In each interval the maximal-degree state reaches $C_{\ell_1,\max}(3)=2$, and the other two lie below.}
\label{fig:circle}
\end{figure*}

\subsection{Two-component cat states}
Let $\ket{c_0}=\ket\alpha$ and $\ket{c_1}=\ket{-\alpha}$, with $s=e^{-2|\alpha|^2}$, and $\ket{\Psi_\pm}=(\ket{c_0}\pm\ket{c_1})/\sqrt{2(1\pm s)}$. Then $|M_{01}|=1/[2(1\pm s)]$, and Eq.~\eqref{eq:l1} gives
\begin{equation}
\El(\Psi_+)=\frac{1-s}{1+s}=\tanh|\alpha|^2,\quad \El(\Psi_-)=1. \label{eq:cat}
\end{equation}
The odd cat state is, up to a global phase, the unique maximal-degree state of two coherent states. Indeed, Eq.~\eqref{eq:max} requires $|x_0|=|x_1|$ and $s\bar x_0x_1$ real, so $x_1=\pm x_0$; the even superposition has $a_j=1+s>1$, which no discrimination attains, and the odd one has $a_j=1-s$ and $S-D_a=s\left(\begin{smallmatrix}1&1\\1&1\end{smallmatrix}\right)\ge0$ for every overlap. In the limit $\alpha\to0$, $\Psi_+$ tends to the vacuum and $\El(\Psi_+)\to0$; $\Psi_-$ tends to the one-photon state and keeps $\El=1$, whereas its $\ell_1$-measure of superposition, $1/(1-s)$, and its robustness of superposition, $\coth|\alpha|^2$~\cite{Torun2026Gram}, diverge (Fig.~\ref{fig:cat}).

\subsection{Three coherent states on a line}
Let $\ket{c_0},\ket{c_1},\ket{c_2}$ be $\ket{-\alpha},\ket0,\ket\alpha$ with $\alpha>0$. With $q=e^{-\alpha^2/2}$ the overlaps are $S_{01}=S_{12}=q$ and $S_{02}=q^4$. By Proposition~\ref{prop:mixture}(b), $S$ is a mixture of groupings iff $2q\le1+q^4$, that is, iff $\alpha$ is at least the threshold amplitude $\alpha_{\rm th}=1.104$, at which $2q=1+q^4$. For the uniform superposition $\ket{\chi}\propto\ket{-\alpha}+\ket0+\ket\alpha$, Theorem~\ref{thm:grouping} then gives
\begin{equation}
\El(\chi)=\frac{6-4q-2q^4}{3+4q+2q^4}\quad(\alpha\ge\alpha_{\rm th}). \label{eq:line}
\end{equation}
For $\alpha<\alpha_{\rm th}$, Theorem~\ref{thm:grouping}(c) shows that $\El(\chi)$ lies strictly below the pairwise bound, and it vanishes with this bound as $\alpha\to0$, where $\chi$ approaches the vacuum, whereas $\Ml(\chi)\to2/3$ (Fig.~\ref{fig:line}).

The three states have a maximal-degree state $\ket\psi$ with alternating signs, $x=(x_0,x_1,x_0)$ and $x_0>0>x_1$. For it, Eq.~\eqref{eq:max} reads
\begin{equation}
x_0\big[(1+q^4)x_0+qx_1\big]=x_1\,(x_1+2qx_0)=\frac13. \label{eq:linemax}
\end{equation}
Equating the two left-hand sides gives
\begin{equation}
\frac{x_1}{x_0}=-\frac{q+\sqrt{4+q^2+4q^4}}{2}. \label{eq:ratio}
\end{equation}
Since $4+q^2+4q^4-9q^2=4(1-q^2)^2>0$, Eq.~\eqref{eq:ratio} gives $|x_1|>2qx_0$, so the second equation in Eq.~\eqref{eq:linemax} fixes $x_0$ for every $\alpha$; since $\sqrt{4+q^2+4q^4}>2-q$, it also gives $|x_1|>x_0$. The condition $S\ge D_a$ also holds. The reflection $\ket{c_0}\leftrightarrow\ket{c_2}$ commutes with $S$ and $D_a$. On the antisymmetric vector $(1,0,-1)$ the condition reads $a_0=1/(3x_0^2)\le1-q^4$; by the first equation in Eq.~\eqref{eq:linemax} this is equivalent to $|x_1|\ge2q^3x_0$, which holds because $|x_1|>2qx_0$ and $q<1$. On the symmetric subspace, $D_a^{-1/2}SD_a^{-1/2}$ has positive off-diagonal elements, so its lower eigenvector has components of opposite signs; this is $D_a^{1/2}x$, with eigenvalue one, so $S\ge D_a$ on this subspace as well. Hence this state is a maximal-degree state for every $\alpha$, with $\El=C_{\ell_1,\max}(3)=2$. Since $|x_1|>x_0$, it is not balanced, and its overlaps $\braket{c_j|\psi}=1/(3\bar x_j)$ differ in modulus, so it is neither golden nor maximally superposed~\cite{Torun2026Gram}. As $\alpha\to0$, $x_1/x_0=-2+\alpha^2+O(\alpha^4)$, and $\ket\psi\propto\ket{-\alpha}+(x_1/x_0)\ket0+\ket\alpha$ tends to the two-photon state $\ket2$.

\subsection{Rotation-symmetric cat states}

Let $\ket{c_j}=\ket{\alpha\omega^j}$, $j=0,\dots,d-1$, with $\omega=e^{2\pi i/d}$: $d$ coherent states on a circle in phase space, with overlaps $S_{jk}=\exp[|\alpha|^2(\omega^{k-j}-1)]$. The Gram matrix is circulant, since the cyclic shift of the reference states is the rotation of the field by $2\pi/d$. The Fourier vectors are eigenvectors of $S$ and define the rotation-symmetric cat states
\begin{equation}
\ket{\Phi_m}\propto\sum_{j=0}^{d-1}\omega^{-jm}\ket{\alpha\omega^j}\propto\sum_{n\equiv m\,(\mathrm{mod}\,d)}\frac{\alpha^n}{\sqrt{n!}}\ket n, \label{eq:circle}
\end{equation}
with the corresponding eigenvalues of $S$
\begin{equation}
\lambda_m=d\,e^{-|\alpha|^2}\!\!\sum_{n\equiv m\,(\mathrm{mod}\,d)}\!\!\frac{|\alpha|^{2n}}{n!},\quad m=0,\dots,d-1. \label{eq:lambda}
\end{equation}
These states include the code words of cat codes, the rotation-symmetric bosonic codes built from coherent states~\cite{Mirrahimi2014,Grimsmo2020}; for $d=2$ they are $\ket{\Psi_\pm}$. Their coefficients $x_j=\omega^{-jm}/\sqrt{d\lambda_m}$ satisfy $Sx=\lambda_mx$, so $\bar x_j\braket{c_j|\Phi_m}=\lambda_m|x_j|^2=1/d$ and $a_j=\lambda_m$ in Eq.~\eqref{eq:max}. The condition $S\ge\lambda_m\id$ holds iff $\lambda_m$ is the smallest eigenvalue. Hence, by Theorem~\ref{thm:max} and in agreement with the optimal unambiguous discrimination of symmetric states~\cite{CheflesBarnett1998}, $\ket{\Phi_m}$ is a maximal-degree state iff $\lambda_m=\lambda_{\min}(S)$; the same states maximize the robustness of superposition~\cite{Torun2026Gram}. For the other states, Eq.~\eqref{eq:sandwich} gives $\EC(\Phi_m)\ge[\lambda_{\min}(S)/\lambda_m]\,\Cmax$.

For every amplitude, therefore, at least one rotation-symmetric cat state is a maximal-degree state: the one with the smallest $\lambda_m$, in which the coherent states interfere most destructively. For small amplitudes, $\lambda_m\approx d|\alpha|^{2m}/m!$, and this state is $\ket{\Phi_{d-1}}$, which tends to the Fock state $\ket{d-1}$. For $d=3$, Eq.~\eqref{eq:lambda} becomes
\begin{equation}
\lambda_m=1+2e^{-3|\alpha|^2/2}\cos\Big(\frac{\sqrt3\,|\alpha|^2}{2}-\frac{2\pi m}{3}\Big),
\end{equation}
so the maximal-degree state changes whenever $\sqrt3|\alpha|^2/2$ passes a multiple of $2\pi/3$: from $\Phi_2$ to $\Phi_0$ at $|\alpha|^2=4\pi/(3\sqrt3)$, that is, $|\alpha|=1.555$, and from $\Phi_0$ to $\Phi_1$ at $|\alpha|=2.199$ [Fig.~\ref{fig:circle}(a)].

These amplitudes have a geometric meaning. Three coherent states at the vertices of an equilateral triangle in phase space have overlaps of equal modulus $s$, and the phase of the Bargmann invariant $S_{01}S_{12}S_{20}$ is twice the oriented area $\mathcal A$ of the triangle. By Corollary~\ref{cor:sym}, the bounds of Eq.~\eqref{eq:bounds} coincide for every state, giving $\El(\rho)=(1-s)\Ml(\rho)$, iff this phase vanishes modulo $2\pi$, that is, iff $\mathcal A$ is an integer multiple of $\pi$. For the inscribed triangle, $\mathcal A=3\sqrt3|\alpha|^2/4$, and this happens exactly at the amplitudes at which the two smallest eigenvalues of $S$ coincide and the maximal-degree state changes. This phase and $s$ also decide whether three reference states with overlaps of equal modulus admit a golden state~\cite{Torun2026Gram}.

\section{Conclusion}\label{sec:conclusion}
Orthogonalization quantifies superposition when it is a measurement rather than a change of basis: the measurements that map every reference state onto a multiple of a basis vector in each outcome are exactly the free operations that convert superposition into coherence. Every coherence measure therefore induces a degree of superposition with its operational meaning, and the measures and tasks of coherence theory become available for superpositions of nonorthogonal states. Because free operations must respect every reference state, the degree depends on the whole reference set.

Distinguishability sets how much superposition can be converted into coherence: an interference term becomes coherence at most with the optimal probability of discriminating its two reference states unambiguously, with equality for two reference states and, for superpositions with positive coefficients of reference states with nonnegative overlaps, exactly when the overlaps are the probabilities of a random grouping. The largest possible degree belongs to the states that yield a maximally coherent state with certainty. They maximize the product of the squared moduli of the coefficients and exist for every set of up to three reference states, whereas the golden states, which are among them, exist for three or more reference states only in special cases.

Every degree bounds the probability of a conversion, and since each coherence measure gives its own degree, two degrees can order two superpositions oppositely and thereby prove them incomparable: neither converts into the other with certainty. Such superpositions exist already for two reference states, whereas all coherence measures order the pure states of a qubit alike, so superposition of two nonorthogonal states is a richer resource than coherence of a qubit. A deterministic conversion that keeps all components, moreover, never lowers the interference term of the norm~\cite{Torun2026Distill}. With the imbalance of the weights, this decides every deterministic conversion between real superpositions of two reference states, in agreement with Gram majorization~\cite{Torun2026Gram}, and for equal positive overlaps it makes constructively interfering superpositions rigid for any number of reference states.

Incomparable superpositions raise the question whether a catalyst, an auxiliary state that is returned unchanged, can enable conversions between them, as catalysts do for entanglement~\cite{JonathanPlenio1999,Datta2023} and coherence~\cite{Bu2016,TakagiShiraishi2022}. Catalytic conversions in RTS remain largely unexplored, and the degrees introduced here, which certify incomparability, are natural tools for studying them.

Unlike the $\ell_1$-measure and the robustness of superposition, which can diverge, the degree stays finite when the reference states become nearly indistinguishable, because it counts only the coherence that can actually be extracted, and at every amplitude it singles out a maximal-degree cat state. It thus provides an operational figure of merit wherever information is carried by superpositions of nonorthogonal states, as in the cat codes of bosonic quantum error correction~\cite{Mirrahimi2014,Grimsmo2020}, which are now stabilized and error-corrected in superconducting circuits~\cite{Ofek2016,Grimm2020}.

\begin{acknowledgments}
Computational tools are a routine part of research, and we acknowledge that the AI system Claude (Anthropic) was used here in the same spirit. In addition, the central idea, how far an orthogonalization can carry superposition into coherence, emerged from an open exchange with it, as did the derivations, numerical checks, figures, and wording. The author verified and organized the content of the paper and takes full responsibility for it.
\end{acknowledgments}

\bibliography{refs}

\end{document}